\documentclass[aps,prb,onecolumn,superscriptaddress,showpacs]{revtex4}
\usepackage{graphicx}
\usepackage{latexsym}
\usepackage{amssymb}
\usepackage{amsmath}
\usepackage{amsfonts}
\usepackage{bm}
\usepackage{multirow}
\usepackage{color}
\usepackage{comment}
\newcommand{\ii}{\mathrm{i}}

\newcommand{\bx}{\mathbf{x}}
\newcommand{\by}{\mathbf{y}}

\newcommand{\he}{\hat{\vec{e}}}

\newcommand{\ha}{\hat{\vec{a}}}

\newcommand{\hb}{\hat{\vec{b}}}
\newcommand{\cE}{\mathcal{E}}
\newcommand{\cB}{\mathcal{B}}
\newcommand{\cA}{\mathcal{A}}
\newcommand{\cH}{\mathcal{H}}
\newcommand{\hE}{\hat{\mathcal{E}}}
\newcommand{\hB}{\hat{\mathcal{B}}}
\newcommand{\hA}{\hat{\mathcal{A}}}

\newcommand{\hn}{\hat{n}}
\newcommand{\htheta}{\hat{\theta}}
\newcommand{\vn}{\vec{\nabla}}

\newcommand{\SO}{\mathrm{SO}}

\newcommand{\SU}{\mathrm{SU}}
\newcommand{\U}{\mathrm{U}}

\newcommand{\beq}{\begin{equation}}
\newcommand{\eeq}{\end{equation}}
\newcommand{\beqn}{\begin{eqnarray}}
\newcommand{\eeqn}{\end{eqnarray}}
\newcommand{\nn}{\nonumber}

\DeclareMathAlphabet{\mathbbold}{U}{bbold}{m}{n}

\def\SU{{\rm SU}}

\def\U{{\rm U}}

\def\dual{{\rm dual}}

\begin{document}

\title{Note on Generalized Symmetries, Gapless Excitations, Generalized Symmetry Protected Topological states, and Anomaly}

\author{Chao-Ming Jian}
\affiliation{Department of Physics, Cornell University, Ithaca,
New York 14853, USA}

\author{Cenke Xu}
\affiliation{Department of Physics, University of California,
Santa Barbara, CA 93106, USA}

\begin{abstract}

We consider quantum many body systems with generalized symmetries,
such as the higher form symmetries introduced recently, and the
``tensor symmetry". We consider a general form of lattice
Hamiltonians which allow a certain level of nonlocality. Based on
the assumption of dual generalized symmetries, we explicitly
construct low energy excited states. We also derive the 't Hooft
anomaly for the general Hamiltonians after ``gauging" the dual
generalized symmetries. A $3d$ system with dual anomalous 1-form
symmetries can be viewed as the boundary of a $4d$ generalized
symmetry protected topological (SPT) state with 1-form symmetries.
We also present a prototype example of $4d$ SPT state with mixed
1-form and 0-form symmetry topological response theory as well as
its physical construction. The boundary of this SPT state can be a
$3d$ anomalous QED state, or an anomalous 1-form symmetry enriched
topological order. Insights are gained by dimensional
compatification/reduction. After dimensional compactification, the
$3d$ system with $N$ pairs of dual 1-form symmetries reduces to a
$1d$ system with $2N$ pairs of dual U(1) global symmetries, which
is the boundary of an ordinary $2d$ SPT state; while the $3d$
system with the tensor symmetry reduces to a $1d$ Lifshitz theory,
which is protected by the center of mass conservation of the
system.

\end{abstract}

\maketitle

\section{Introduction}

Various lattice models with different emergent gauge invariance
were constructed in the context of quantum many-body condensed
matter systems, including models with emergent $\U(1)$ gauge
invariance~\cite{wen2003,ms2003,hermele2004}, and models with more
exotic tensor like gauge
transformations~\cite{xu2006a,gu2006,xu2006b,xu2008,xuhorava,xu2016}.
The most well-known example is the quantum spin ice system with
emergent electromagnetism and photon like excitations at low
energy, as well as Dirac monopoles~\cite{spinice}. The analysis of
these lattice models usually relies on the ``spin-wave" expansion,
meaning expanding the theory at certain presumed semiclassical
mean field minimum of the Hamiltonian, or saddle point of the
action in path integral. A low energy field theory is derived from
this procedure (for example the Maxwell theory), then it is
expected that this field theory captures the infrared physics of
the lattice model at long scale. The stability of the state of
interests described by the low energy field theory usually needs
to be studied case-by-case for each particular example. The
general procedure of such analysis is that, one treats the
deviation from the field theory as perturbations, and demonstrate
that these perturbations are irrelevant under renormalization
group flow at the desired state described by the field theory. But
for a general form of lattice Hamiltonian, it is unclear whether
such a mean-field minimum (and its corresponding field theory)
really exists, or whether the perturbative renormalization group
argument is reliable because the deviation from the desired state
can be too strong to be treated perturbatively. For example, it is
known that the lattice model for the emergent photon phase can be
tuned into states very different from the ordinary Maxwell theory,
such as the confined phase with various spin or valence bond solid
orders, and the Rokhsar-Kivelson (RK) point with nonrelativistic
dispersion~\cite{rk,ms2003,fradkin2004}.

Sometimes the argument for the stability of the desired low energy
state can also be translated to certain physical picture, for
example the behavior of topological defects such as the Dirac
monopoles; namely depending on whether the Dirac monopoles are
gapped or condensed, the lattice gauge theory is in its deconfined
or confined phases. But this argument relies heavily on the
specific form of the theory, since the physical picture and the
theory describing the condensation of topological defects can vary
significantly between lattice theories with different generalized
gauge transformations~\cite{xu2016}.

Recently new tools and languages such as generalized higher-form
symmetries were introduced to analyze gauge
fields~\cite{formsym0,formsym1,formsym2,formsym3,formsym4,formsym5,formsym6,formsym7,formsym8},
and various features of gauge fields such as the physical
consequence of a topological term can be clearly studied following
this language~\cite{formanomaly}. In the current manuscript, the
most fundamental assumption we make about the systems under study
is that, though our system is defined on a lattice, at least at
the long scale there exists a $\U(1)^g$ symmetry. $\U(1)^g$ is a
generalized $\U(1)$ symmetry such as the higher-form symmetry or a
``tensor symmetry", whose definition will be spelled out later in
the manuscript. The $\U(1)$ nature of the symmetry means that the
charges of the generalized symmetry take arbitrary discrete
integer eigenvalues, and charges with different supports in space
all commute with each other. $\U(1)^g$ can be an actual symmetry
on the lattice scale (UV scale), it can also be of emergent
nature, meaning it only exists at long scale.

Depending on the dimensionality, there exists a topological
soliton associated with this presumed $\U(1)^g$ symmetry. The
topological soliton is defined in space but not space-time, and it
has a smooth spatial energy distribution without singularity (for
example, a Dirac monopole is considered as a defect, instead of
soliton). We then further assume that at long scale the
topologically quantized soliton number is conserved, which means
that the system also has an emergent $\U(1)^g_{\dual}$ symmetry.
Hence at the long scale, there exists a dual structure with an
enlarged $\U(1)^g \times \U(1)^g_{\dual}$ symmetry where the two
$\U(1)^g$ and $\U(1)^g_{\dual}$ symmetries act on two sets of
degrees of freedom that are related to each other in a non-local
way. 
In this work, we will discuss the physical implications of the
presumed infrared $\U(1)^g \times \U(1)^g_{\dual}$ symmetry of
general lattice Hamiltonians, without relying on expansion at
classical saddle point.

This manuscript is arranged as follows: in section II, we start
with the assumption of the dual generalized symmetries in the
infrared, then demonstrate the 't Hooft anomaly and a stable
gapless phase on general grounds, without assuming any space-time
symmetry or semiclassical treatment of the lattice model; in
section III, we identify the gapless phases discussed in section
II as the boundary of higher dimensional symmetry protected
topological (SPT) states with 1-form symmetries, and make
connection to ordinary SPT states after dimensional
compactification/reduction; section IV introduces a prototype
example of $4d$ SPT state with both 1-form and 0-form symmetries,
whose boundary is a prototype example of anomalous 1-form symmetry
enriched topological states (1-form SET) with fractionalization of
1-form symmetry; in section V we generalize the discussions in
previous section to the case with ``tensor gauge transformation",
meaning we derive the 't Hooft anomaly of a pair of dual tensor
symmetries, and demonstrate the gaplessness of the spectrum based
on the assumption of dual tensor symmetries; we also show that
after dimensional compactification the system with a pair of dual
tensor symmetries reduces to a $1d$ Lifshitz theory.

\section{3d systems with $\U(1)$ 1-form symmetry}

\label{Sec3d_U(1)_1form}

\subsection{Consequences of 1-form symmetries}

For our purpose, we do not take a specific example of state of
matter, and show that this example has a 1-form symmetry. Instead,
we start with the assumption that at least at the long scale, our
$3d$ system has a $\U(1)^g$ symmetry, where $\U(1)^g$ is a 1-form
symmetry~\cite{formsym0,formsym1,formsym2,formsym3,formsym4,formsym5,formsym6,formsym7,formsym8}.
We will explore what consequences this general assumption can lead
to. Here $3d$ means 3 spatial dimensions.

There is a 1-form charge density associated with this presumed
$\U(1)^g$ symmetry: $Q_{A} = \int^A d\vec{S} \cdot \vec{\rho}$.
The integral is over a two dimensional surface $A$. The
conservation of the charge density means that the 1-form charge
cannot be created or annihilated, but it can ``leak" through the
boundary of $A$ through a 1-form symmetry current. But if $A$ is a
closed surface without any boundary, $Q_{A}$ must be a constant,
namely \beqn Q_{A} = \int^{A}_{\partial A =\emptyset} d\vec{S}
\cdot \vec{\rho} = \int^V_{\partial V = A} d^3x \ \vn \cdot
\vec{\rho} = \mathrm{const}.  \eeqn Since this must be valid for
any closed surface, it implies that $\vn \cdot \vec{\rho}$ is a
time-independent constant everywhere in the entire space at long
scale. Hence $\vec{\rho}$ can be viewed as an electric field
$\vec{e}$ which satisfies the Gauss law constraint. The equation
of motion of the ordinary electromagnetic field, $i.e.$ the
Maxwell equations, can be viewed as the continuity equation of the
1-form symmetries: \beqn
\partial_\mu J^{(e)}_\mu = \frac{\partial e_i}{\partial t} -
\partial_j \epsilon_{ijk} b_k = 0, \cr\cr \partial_\mu J^{(m)}_\mu =
\frac{\partial b_i}{\partial t} + \partial_j \epsilon_{ijk} e_k =
0. \eeqn This means that for the ordinary Maxwell theory, the
currents of the two 1-form symmetries are: \beqn && J^{(e)} =
(\rho^{(e)}_i, \ J^{(e)}_{ij}) = (e_i, \ \epsilon_{ijk} b_k),
\cr\cr && J^{(m)} = (\rho^{(m)}_i, \ J^{(m)}_{ij}) = (b_i, \
-\epsilon_{ijk} e_k). \label{current}\eeqn This is analogous to
the more familiar fact that, the equation of motion of a
superfluid is also the continuity equation of its super-current.
Note that the conserved current $J^{(e)}$ is associated with the
aforementioned 1-form $\U(1)^g$ symmetry. The conserved current
$J^{(m)}$ will be associated with a different 1-form symmetry,
denoted as $\U(1)^g_{\rm dual}$, whose physical meaning and
definition will be explained later in the section.

Let us denote the operator of the electric field as $\he$. When a
quantized electric field is realized in condensed matter systems,
it usually only takes discrete integer eigenvalues, because the
physical meaning of the electric field operator is usually the
number operator of certain quantum boson (for example the dimer
number operator~\cite{qd,ms2003}), or spin component
$S^z$~\cite{wen2003,hermele2004}. We consider a lattice model for
these electric field operators like previous literatures on
quantum spin ices. If the Gauss law constraint is imposed strictly
on the lattice, the 1-form symmetry is a microscopic symmetry of
the system. However in condensed matter realizations the Gauss law
constraint is usually not imposed strictly on the lattice, instead
there is a large energy penalty for creating defects that violate
the Gauss law constraint (for example in the quantum spin ice, the
ice-rule of the spin configuration was enforced energetically).
The Gauss law constraint and hence the 1-form symmetry (now we
refer to it as the electric 1-form symmetry) is only an emergent
symmetry at long scale.

Using the (emergent) Gauss law constraint, one can prove that the
Hamiltonian of the system {\it must} have a gauge invariance:
$\ha(\bx) \rightarrow \ha(\bx) + \vn f$, where $\hat{a}_i =
\hat{a}_i + 2\pi$ is the canonical conjugate operator of
$-\hat{e}_i$, i.e. $[\hat{e}_i(\bx), \hat{a}_{i'}(\bx')]=\ii
\delta_{ii'} \delta(\bx-\bx')$. Here, we've chosen the convention
that $-\hat{e}_i$ is the canonical conjugate momentum of
$\hat{a}_i$ to match the convention of the ordinary Maxwell
theory. We will defer the proof to the example with the ``tensor
symmetry" (section V) we will discuss. Here we state that by
assuming there is a $\U(1)^g$ 1-form symmetry, the Hamiltonian of
the system must have a $\U(1)$ gauge invariance. Hence a local
Hamiltonian of the system will only involve gauge invariant
operators such as $\he$ and $\hb = \vn \times \ha$. Generally, a
local Hamiltonian of the system that respects the 1-form $\U(1)^g$
symmetry takes the form: \beqn H = \sum_{\bx} \ \cH[\he(\bx), \
\hb(\bx)] \label{Hem}\eeqn $\cH[X, Y]$ must be a periodic function
of $Y$, because $\hb(\bx)$ and $\ha(\bx)$ are both periodically
defined at any spatial location $\bx$. This means that a $2\pi$
flux has no physical effect if it is only inserted through a
single plaquette of the lattice. The flux only affects physics
when it is spread out in space, hence there are nontrivial fluxes
through plaquettes which are not multiple of $2\pi$. We do not
assume any space-time symmetry in $H$, hence $\cH$ can involve
mixture terms such as $\hat{e}(\bx)_i^n \sin ( \hat{b}(\bx)_j)^m +
H.c.$. $\cH$ also does not need to be translationally invariant,
$i.e.$ it can have disorder. Here, we mainly focus on the local
Hamiltonian of the form Eq.~\ref{Hem}. But our analysis on the
local Hamiltonian Eq.~\ref{Hem} can be extended to systems with
certain degree of non-locality.

Now we are ready to define the dual $\U(1)^g_\dual$ symmetry.
Since $\hb = \vn \times \ha$, it appears that the magnetic charge
density vanishes $\vn \cdot \hb = 0$. But just like the existence
of vortices in superfluid, there exists singular defects like
Dirac monopoles which complicate the scenario. We assume that $\vn
\cdot \hb = 0$ holds at low energy or long scale, hence
$\int^A_{\partial A = \emptyset} d\vec{S} \cdot \hb = 0$ for a
large enough closed surface $A$ (unless $A$ has nontrivial winding
over the entire space), $i.e.$ there is a $\U(1)^g \times
\U(1)^g_\dual$ 1-form symmetry at long scale. This is similar to
the physical picture that the Dirac monopole defect has a large
energy gap, hence positive and negative monopole pairs must be
tightly bound at low energy. For the ordinary Maxwell theory, the
current associated to the $ \U(1)^g_\dual$ symmetry is given by
the second line of Eq.~\ref{current}. In the following, we will
discuss the consequence of the $\U(1)^g \times \U(1)^g_\dual$
symmetry in the general Hamiltonian Eq.~\ref{Hem} of which the
ordinary Maxwell theory is only a special case.

For a general Hamiltonian given in Eq.~\ref{Hem}, using the
Heisenberg equation, we can derive the 1-form currents for both
the electric and magnetic 1-form symmetries: \beqn \frac{\partial
\hat{e}_i(\bx)}{\partial t} &=& \ii [H, \ \hat{e}_i(\bx)] =
 \int d\by \  \ii \frac{\partial
\cH}{\partial \hat{b}_k(\by)} \epsilon_{ji'k}\partial_{y_j}
[\hat{a}_{i'}(\by), \ \hat{e}_i(\bx) ] =
\epsilon_{ijk}\partial_{x_j} \frac{\partial \cH}{\partial
\hat{b}_k(\bx)}, \cr \cr \frac{\partial \hat{b}_i(\bx)}{\partial
t} &=& \ii [H, \ \hat{b}_i(\bx)] =
 \int d\by \  \ii \frac{\partial
\cH}{\partial \hat{e}_{k'}(\by)} \epsilon_{ijk}\partial_{x_j}
[\hat{e}_{k'}(\by) ,\ \hat{a}_{k}(\bx) ] = -
\epsilon_{ijk}\partial_{x_j} \frac{\partial \cH}{\partial
\hat{e}_k(\bx)} \label{general_current} , \eeqn which can be
viewed as the generalized 1-form electric and 1-form magnetic
current conservation equations. The charges associated to 1-form
electric and 1-form magnetic symmetries are still identified as
$\he$ and $\hb$. The 1-form symmetry currents for a general
Hamiltonian are \beqn J^{(e)}_{ij}(\bx) = \epsilon_{ijk}
\frac{\partial \cH}{\partial \hat{b}^k(\bx)}, \ \ \
J^{(m)}_{ij}(\bx) = - \epsilon_{ijk} \frac{\partial \cH}{\partial
\hat{e}^k(\bx)} \label{emCurrent_Def}\eeqn respectively.

The $\U(1)^g \times \U(1)^g_{\dual}$ dual 1-form symmetries have
the 't Hooft anomaly. For the ordinary Maxwell theory, this
anomaly can be seen by the form of the 1-form currents
Eq.~\ref{current}: the current of $\U(1)^g$ symmetry is the charge
density of the $\U(1)^g_\dual$ symmetry, and vice versa. This
means that the process of generating a current associated to one
symmetry, necessarily violates the conservation of the charge of
the other symmetry. Hence there must be a mixed anomaly between
these two symmetries. The mixed $\U(1)^g \times \U(1)^g_{\dual}$
anomaly of the ordinary (3+1)$d$ Maxwell theory was derived in
previous literatures such as Ref.~\onlinecite{Cordova2019}.

In the following, we derive the 't Hooft anomaly for systems
described by the general Hamiltonian Eq.~\ref{Hem}, which has the
$\U(1)^g \times \U(1)^g_{\dual}$ 1-form symmetries. To demonstrate
the anomaly, we start by gauging the 1-form symmetries, i.e. by
coupling $J^{(e)}$ and $J^{(m)}$ to external gauge fields
$A^{(e)}$ and $A^{(m)}$, both of which are rank-2 tensor (2-form)
gauge fields. $A^{(e)}$ and $A^{(m)}$ carry with them the
following gauge transformations: \beqn && A^{(e,m)}_{i,0}
\rightarrow A^{(e,m)}_{i,0} +
\partial_t f^{(e,m)}_i,  \cr \cr && A^{(e,m)}_{ij} \rightarrow
A^{(e,m)}_{ij} + \partial_j f^{(e,m)}_i - \partial_i f^{(e,m)}_j.
\label{em2formGaugeTrans} \eeqn These tensor gauge fields are
antisymmetric: $A_{ij}^{(e,m)} = -A_{ji}^{(e,m)}$.

To explain how the rank-2 tensor gauge fields $A^{(e,m)}$ couple
to the system described in Eq. \ref{Hem}, we need to switch to a
Lagrangian formalism of the problem. Before turning on the gauge
fields $A^{(e,m)}$, the Lagrangian of the system is given by
\begin{align}
\mathcal{L} = \sum_{\bf x} e_i (\bx ) \frac{\delta H}{\delta
e_i(\bx)} - \cH[\vec{e}(\bx), \ \vec{b}(\bx)],
\end{align}
where $\vec{e} (\bx ) $ and $\vec{b}(\bx)$ should be viewed as
fields (instead of as operators). In the Legendre transformation,
$\dot{a}_i (\bx ) = - \delta H/ \delta e_i (\bx)$, which allows us
to express $\vec{e} (\bx)$ as a function of $\dot{\vec{a}} (\bx )$
and $\vec{b}(\bx)$, and further, to write the Lagrangian as a
function of $\dot{\vec{a}} (\bx )$ and $\vec{b}(\bx)$, namely
$\mathcal{L}[\dot{\vec{a}} (\bx ), \vec{b}(\bx )]$. Under the
electric 2-form gauge transformation (whose action on $A^{(e)}$
are given in Eq.~\ref{em2formGaugeTrans}), the degrees of freedom
in the Lagrangian $\mathcal{L}[\dot{\vec{a}} (\bx ), \vec{b}(\bx
)]$ transform as
\begin{align}
& a_i \rightarrow a_i - f^{(e)}_i,
\nonumber \\
& \dot{a}_i \rightarrow \dot{a}_i - \partial_t f^{(e)}_i,
\label{e2formGaugeTrans}
\\
& b_i \rightarrow b_i - \epsilon_{ijk}\partial_j f^{(e)}_k.
\nonumber
\end{align}
When the system is coupled to the background two-form gauge fields
$A^{(e,m)}$, it can be described by the Lagrangian
\begin{align}
\mathcal{L}_g & = \mathcal{L}\left[\dot{a}_i + A^{(e)}_{i,0} , b_i
- \frac{1}{2}\epsilon_{ijk} A^{(e)}_{jk} \right] \nonumber
\\
& ~~~~ + \sum_{\bx} \frac{1}{2\pi} \left( A_{ij}^{(m)} (\bx)
J_{ij}^{(m)} (\bx ) + A_{i,0}^{(m)} (\bx) b_i(\bx) \right)
\nonumber  \\
& = \mathcal{L}\left[\dot{a}_i + A^{(e)}_{i,0} , b_i -
\frac{1}{2}\epsilon_{ijk} A^{(e)}_{jk} \right] \nonumber
\\
& ~~~~ + \sum_{\bx} \frac{1}{2\pi} \left( -A_{ij}^{(m)} (\bx)
\epsilon_{ijk} \dot{a}_k(\bx ) + A_{i,0}^{(m)} (\bx) b_i(\bx)
\right)
\end{align}
One can easily check that, when $A^{(m)} =0$, the Lagrangian
$\mathcal{L}_g$ is invariant under the electric 2-form gauge
transformations given by  Eq. \ref{em2formGaugeTrans} and Eq.
\ref{e2formGaugeTrans}. The coupling to the magnetic 2-form gauge
field $A^{(m)}$ is introduced in $\mathcal{L}_g$ in the form of
minimal coupling. Here, we have made use of the general definition
of $J_{ij}^{(m)} $ given in Eq.~\ref{emCurrent_Def} as well as the
fact that $\dot{a}_i (\bx ) = - \delta H/ \delta e_i (\bx)$.

It turns out that, when $A^{(m)} \neq 0$, the Lagrangian
$\mathcal{L}_g$ is no longer invariant under the electric 2-form
gauge transformation:
\begin{align}
\mathcal{L}_g \rightarrow \mathcal{L}_g + \sum_{\bx}
\frac{1}{2\pi} \left( A_{ij}^{(m)}  \epsilon_{ijk} \partial_t
f^{(e)}_k - A_{i,0}^{(m)} \epsilon_{ijk} \partial_j f^{(e)}_k
\right),
\end{align}
which indicates a mixed 't Hooft anomaly of the U$(1)^g \times
\U(1)_{\rm dual}^g$ symmetry in the system. In fact, this anomaly
matches that of the boundary theory of a $(4+1)d$
symmetry-protected topological (SPT) state that has the 1-form
U$(1)^g \times \U(1)_{\rm dual}^g$ symmetry and a topological
response given by~\cite{mcgreevy,formsym6,Cordova2019} \beqn
\mathcal{S}_{\mathrm{CS}} = \int d\tau d^4x \ \frac{1}{2\pi}
A^{(e)} \wedge d A^{(m)} \label{CS}\eeqn Hence if the $\U(1)^g
\times \U(1)^g_{\dual}$ symmetries are microscopic symmetries, the
$3d$ state described by the Hamiltonian Eq.~\ref{Hem} must be a
boundary state of a $4d$ SPT state with 1-form symmetries. Here,
$A^{(e)}$ and $A^{(m)}$ are treated as two-form fields in
$(4+1)d$.

\subsection{Gapless excitations of systems with dual 1-form symmetries}

Common wisdom says that a mixed 't Hooft anomaly of the dual
$\U(1)^g \times \U(1)^g_\dual$ symmetry implies that the spectrum
of the $3d$ system cannot be trivially gapped, namely the
Hamiltonian $H$ cannot have a unique ground state and gapped
spectrum in the thermodynamics limit. Here we explicitly construct
an excited state of the general Hamiltonian Eq.~\ref{Hem} with
vanishing energy in the thermodynamics limit. We define our system
on a three dimensional cubic lattice which forms torus with size
$L^3$, and we assume there is a unique ground state of $H$ denoted
by $|\Omega \rangle$. We consider the following state
$|\Psi\rangle$: \beqn |\Psi\rangle = \hat{O}_{q} |\Omega\rangle, \
\ \ \hat{O}_{q} = \exp\left( \ii q \sum_{\bx} x \frac{2\pi
\hat{e}_y(\bx)}{L^2} \right), \label{Oq1} \eeqn where
$\hat{O}_{q}$ is a function of $\he$ only, and it creates a
magnetic flux quantum $2\pi q$ with size $L^2$ along the $\hat{z}$
direction. $\hat{O}_q$ shifts $\hat{a}_y$ by $\hat{a}_y
\rightarrow \hat{a}_y + 2\pi x/L^2$. Hence the gauge invariant
Wilson loop $W_y = \exp(\ii \int_0^L dy \hat{a}_y)$ still has a
periodic boundary condition after the shift, $i.e.$ $W_y(x = 0) =
W_y(x = L)$ for integer $q$. Notice that since $\hat{O}_{q}$ is a
function of $\he$, $\hat{O}_{q}$ must commute with any composite
operator of $\he$. This operator inserts flux $2\pi q/L^2$ on
every plaquette in the XY plane. Using the language in
Ref.~\onlinecite{hermele2004}, The state $|\Psi \rangle$ carries a
nontrivial topological charge; but using more recently developed
language, $|\Psi \rangle$ carries a different 1-form $\U(1)^g_{\rm
dual}$ symmetry charge compared with the ground state. To be more
precise, this symmetry charge here is $\int dx dy \ \hat{b}_z$
(with the integration over the XY-plane).

Since we made a powerful assumption that there is an emergent
magnetic 1-form symmetry $\U(1)^g_\dual$ at long scale, the
assumption of $|\Omega\rangle$ being the unique ground state
implies that it is also an eigenstate of the 1-form
$\U(1)^g_\dual$ charges. $|\Psi\rangle$ must be orthogonal to
$|\Omega \rangle$ when the size of the created soliton is large
compared with the lattice constant, because these two states carry
different charges under $\U(1)^g_\dual$. Though $|\Psi\rangle$ is
not necessarily the eigenstate of the Hamiltonian, the energy of
$|\Psi\rangle$ is evaluated as \beqn && E_\Psi = \langle \Psi | H
| \Psi \rangle = \langle \Omega | \hat{O}^\dagger_{q} H
\hat{O}_{q} |\Omega\rangle \cr\cr &=& \sum_{\mathbf{x}} \langle
\Omega | \cH[\he(\bx), \ \hb(\bx) + \frac{2\pi q}{L^2} \hat{z} ]
|\Omega\rangle \cr\cr &=& E_{\Omega} + \sum_{\mathbf{x}} \sum_{m =
1}^\infty \frac{1}{m!} \langle \Omega |
\partial^m_{\hat{b}_z}\cH[\he(\bx), \ \hb(\bx)] |\Omega\rangle
\left(\frac{2\pi q}{L^2} \right)^m, \eeqn where $\hat{z}$ is the
unit vector along the $z$ direction. We have expanded the
energy as a polynomial of $1/L^2$. For our purpose we only need to
worry about the leading order expansion of $1/L^2$, because all
the other terms will vanish under the limit $L \rightarrow
\infty$.

The leading order expansion of $E_\Psi$ involves the following
terms: \beqn \sum_{\bx} \langle \Omega |
\partial_{\hat{b}_z}\cH[\he(\bx), \ \hb(\bx)]| \Omega \rangle
\frac{2\pi q}{L^2}. \label{ex}\eeqn For a general state this
expectation value does not vanish. However, since $|\Omega
\rangle$ is the ground state, $ \langle \Omega | \sum_{\bx}
\partial_{\hat{b}_z} \cH[\he(\bx), \ \hb(\bx)]| \Omega
\rangle$ must vanish because otherwise one can always choose the
sign of $q$ to make the energy of $|\Psi\rangle$ lower than
$|\Omega\rangle$ for large enough $L$, which violates the
assumption that $|\Omega\rangle$ is the ground state.

Let us review our logic here: we do not first take the ordinary
Maxwell theory and demonstrate that there is a 1-form symmetry;
instead we start with the assumption that there exists one 1-form
symmetry $\U(1)^g$ at long scale, then demonstrated that there
must be a gauge invariance as a consequence of the 1-form
symmetry. And the gauge invariance allows us to define the dual
1-form symmetry $\U(1)^g_\dual$. Then by further assuming $\U(1)^g
\times \U(1)^g_\dual$ at long scale, we constructed a state that
is orthogonal to the ground state, with energy approaching the
ground state in the thermodynamics limit. The construction also
does not rely on the semiclassical ``spin-wave" expansion used
often in literature of lattice quantum spin or boson models.
Similar ``soliton insertion" argument was used in the original
Lieb-Shultz-Matthis theorem~\cite{lsm}, and the Luttinger
theorem~\cite{oshikawa}.

The argument above can go through even with a certain degree of
non-locality in the Hamiltonian. For example, if there is a term
in the Hamiltonian \beqn H' = \sum_{\bx, \bx'} f(|\bx - \bx'|)
F[\hb(\bx)]F[\hb(\bx')], \eeqn one can show that as long as
$f(|\bx|)$ falls off faster than $1/|\bx|^2$ at the long distance,
the state $|\Psi\rangle$ constructed above still has vanishing
energy with $L \rightarrow \infty$.

\section{Generalized SPT states and dimensional reduction}

Helpful further insights can be gained through compactifying the
system discussed in the previous section to one dimension. The
mixed 't Hooft anomaly between the two dual 1-form symmetries will
reduce to a mixed anomaly of two ordinary (0-form) $\U(1)$
symmetries. The $4d$ bulk will reduce to a $2d$ bosonic SPT state
with ordinary (0-form) symmetries.

We compactify the YZ plane to a $2d$ torus with a small size.
Since the $1d$ system is along the $\hat{x}$ direction, a $2d$
surface $A$ wrapping around the $1d$ line could be either in the
XY plane, or the XZ plane. In the $3d$ systems with the $\U(1)^g
\times \U(1)^g_\dual$ 1-form symmetry, there is a 1-form charge
associated with the compactified XZ plane: \beqn \int_{\bx \in
\mathrm{XZ}} d^2x \ \hat{e}_y(\bx) \sim \int dx \ \hn(x). \eeqn
Since the system is highly compact in the Y and Z directions, we
ignore the modes with finite discrete momenta in these directions.
In other words, all the fields are constants in these two
directions. Then, we can define a $1d$ particle density $\hn(x)
\sim \hat{e}_y(x) $ in this compactified system. After proper
normalization, we can also define the canonical conjugate variable
of $\hn(x)$, $i.e.$ the phase angle operator $\htheta(x)$ as \beqn
\int_{\bx \in \mathrm{XY}} d^2x \ \hat{b}_z(\bx) \sim \int dx \
\nabla_x \htheta(x), \eeqn $\htheta(x) \sim \hat{a}_y(x)$.
$\htheta(x) $ and $\hn(x)$ obey the standard commutation relation:
$[\htheta(x), \ \hn(x')] = - \ii \delta_{x,x'}$. The 1-form
symmetries discussed in previous examples becomes the ordinary
global symmetries (0-form symmetries) in $1d$.

The $\U(1)^g_\dual$ charge now becomes the topological soliton
number in this $1d$ system: \beqn N_T = \frac{1}{2\pi}\int_0^L dx
\ \nabla_x \htheta(x). \eeqn The general Hamiltonian we considered
in Eq.~\ref{Hem} becomes a $1d$ Hamiltonian with an ordinary
$\U(1)$ symmetry \beqn H = \sum_x \ \cH[\hn(x), \nabla_x
\htheta(x)]. \eeqn

All the analysis in Sec. \ref{Sec3d_U(1)_1form} has counterparts
in the compactified system. We assume that at long scales both the
particle number $\int dx \ \hn(x)$ and the topological soliton
number $N_T$ are conserved, namely there is a $\U(1) \times
\U(1)_\dual$ symmetry at long scale. We denote the ground state of
the Hamiltonian described above as $|\Omega\rangle$, and then
consider the following state $|\Psi \rangle$: \beqn |\Psi\rangle =
\hat{O}_q|\Omega\rangle = \exp\left( \ii q \sum_x \ \frac{2\pi
\hn(x)}{L} x \right) | \Omega \rangle. \eeqn The operator
$\hat{O}_q$ is the analogue of the operator $\hat{O}_q$ in
Eq.~\ref{Oq1} compactified to $1d$. With $q=1$, $|\Psi \rangle$
contains one extra soliton $\hat{N}_T$ compared with the ground
state $|\Omega \rangle$: $\hat{O}_1$ creates one extra winding of
$\hat{\theta}$ in the $1d$ system. Since we've assumed that the
$\U(1)_\dual$ is an emergent symmetry at long scale,
$|\Psi\rangle$ must be orthogonal to the ground state. The
evaluation of the energy of $|\Psi\rangle$ is similar to the
discussion in Sec.~\ref{Sec3d_U(1)_1form}. We can show that the
energy of $|\Psi\rangle$ approaches the energy of $|\Omega
\rangle$ as $L \rightarrow \infty$.

When the system is reduced to $1d$, its $\U(1) \times \U(1)_\dual$
symmetry has an ordinary 't Hooft anomaly. In fact, the action of
the $\U(1) \times \U(1)_\dual$ in the reduced $1d$ system mimics
the spin and charge U(1) symmetry action on the boundary of a $2d$
quantum spin Hall insulator. It is known that the boundary of the
quantum spin Hall insulator with both charge and spin U(1)
symmetries has a mixed perturbative 't Hooft anomaly. To show this
anomaly formally, one can couple the charge $\U(1)$ current to a
$\U(1)^{(e)}$ back ground gauge field $A^{(e)}$, and couple the
spin $\U(1)$ (or the $\U(1)_\dual$) current to another background
$\U(1)^{(m)}$ gauge field $A^{(m)}$. This mixed anomaly is
identical to the boundary of a $(2+1)d$ bulk Chern-Simons theory
\beqn \mathcal{S} = \int d\tau d^2x \ \frac{1}{2\pi} A^{(e)}
\wedge d A^{(m)}. \label{anomaly1d1} \eeqn Physically, this
anomaly simply means that the current of one $\U(1)$ symmetry is
the charge density of the other $\U(1)$ symmetry, hence a process
of creating the current of one $\U(1)$ symmetry would necessarily
violate the conservation of the charge of the other $\U(1)$
symmetry.

There is another pair of dual $\U(1)$ symmetries in the $1d$
system after compactification, which originates from the $3d$ dual
$\U(1)$ 1-form symmetries: the U(1) symmetry generated by
$\int_{\bx \in \mathrm{XY}} d^2x \ \hat{e}_z(\bx)$, and the
$\U(1)_{\dual}$ symmetry associated to the conservation of
$\int_{\bx \in \mathrm{XZ}} d^2x \ \hat{b}_y(\bx)$. There is also
a mixed 't Hooft anomaly between these two dual $\U(1)$
symmetries. Hence one pair of dual 1-form symmetries in $3d$ will
reduce to two pairs of ordinary dual symmetries in $1$d. In
general, if we start with $N$ pairs of dual $\U(1)^g \times
\U(1)^g_\dual$ 1-form symmetries in $3d$, after compactification
to $1d$ there will be $2N$ pairs of dual $\U(1) \times
\U(1)_\dual$ symmetries in $1d$. The $4d$ bulk system for the $3d$
system with a series of 1-form symmetries can have a Chern-Simons
response theory \beqn \mathcal{S} = \int d\tau d^4x \
\frac{1}{4\pi} K_{IJ} C^I \wedge d C^J, \eeqn where $C^I$ is a two
form gauge field, and $K^{IJ}$ is an antisymmetric matrix. Then
after dimensional reduction as discussed in this section, the
corresponding $2d$ bulk theory for the $1d$ system should have a
CS response theory \beqn \mathcal{S} = \int d\tau d^2x \
\frac{1}{4\pi} K'_{IJ} C^I \wedge d C^J, \ \ \ K' = K \otimes
\left(
\begin{array}{cc}
0 & -1 \\
1 & 0
\end{array}
\right). \eeqn In the $(2+1)d$ system $C^I$ is a 1-form gauge
field, and $K'$ is a symmetric matrix. Hence the $4d$ generalized
SPT state can be studied and understood as its $2d$ counterpart
with ordinary symmetries after dimensional reduction.


\section{A prototype SPT state with mixed 0-form and 1-form symmetries, 1-form Symmetry Enriched Topological Order}

Eq.~\ref{CS} is a $(4+1)d$ topological response theory involving
only 1-form symmetries. In general, if there is an extra ordinary
(0-form) symmetry $G$ in the system, one can also consider the
mixed topological response theory between the 0-form symmetry $G$
and the 1-form symmetries. For example, we can consider a $(4+1)d$
bulk system which has a topological response \beqn
\mathcal{S}_{\mathrm{topo}} = \pi \int d\tau d^4x \
w_2[A^{\SO(3)}] \cup \frac{d A^{(e)}}{2\pi} . \label{response}
\eeqn Here, $A^{\SO(3)}$ is the background (1-form) gauge field
associate to the 0-form symmetry $G=\SO(3)$ and $w_2$ is the
second Stiefel-Whitney class.

A candidate system with this response theory can be constructed as
follows: we start with a $(4+1)d$ QED with a microscopic electric
$\U(1)^g$ 1-form symmetry. We will denote the $(4+1)d$ bulk
dynamical gauge field as $a_\mu$. There is no microscopic magnetic
higher-form symmetry, hence there are defects with their own
dynamics analogous to the Dirac monopole. The Dirac monopole
defect in $(4+1)d$ is a one dimensional line/loop. It was shown
that the ``decorated defect" construction is a very powerful
physical picture of constructing SPT states with 0-form
symmetries~\cite{senthilashvin,chenluashvin}, i.e. higher
dimensional SPT states can be constructed by decorating the
topological defects of order parameters with lower dimensional SPT
states. Here we also follow the recipe of decorated defects: we
attach the Dirac monopole line in $(4+1)d$ bulk with a one
dimensional ordinary SPT phase with $G = \SO(3)$ symmetry, i.e.
the Haldane phase, and proliferate the Dirac monopole line. The
$(4+1)d$ bulk will be driven into a gapped and confined phase,
while the most natural $(3+1)d$ boundary state of the system will
be a QED whose Dirac monopole carries a spin-1/2 under the 0-form
$\SO(3)$ symmetry, while there is no electric charge. A theory
which describes this boundary state is the $(3+1)d$ CP$^1$ model:
\beqn \mathcal{S}_{\mathrm{boundary}} = \int d\tau d^3x \
\sum_{\alpha = 1}^2 |(\partial - \ii \tilde{a})z_\alpha |^2 +
\cdots \label{cp1} \eeqn where $z_\alpha$ represents a spin-1/2
representation of the $\SO(3)$ 0-form symmetry carried by the
boundary termination of the Dirac monopole line in the $(4+1)d$
bulk, while $\tilde{a}_\mu$ is the ``dual" gauge field of $a_\mu$
at the $(3+1)d$ boundary whose gauge charge is the Dirac monopole
of $a_\mu$. As we can see from its topological response
$\mathcal{S}_{\mathrm{topo}}$, this $(4+1)d$ bulk is an SPT state
protected by the electric 1-form U(1) symmetry and the 0-form
symmetry $G$. Its boundary state cannot be gapped with a unique
ground state without breaking the symmetries. One way to
understand it is to consider the compactification of 3 spatial
dimensions to a 3-dimensional sphere $S^3$ with a non-trivial flux
$\int_{S^3} dA^{(e)} = 2\pi$. The effective $(1+1)d$ system after
the dimensional compactification/reduction has a topological
response identical to the SO(3) symmetric Haldane phase in
$(1+1)d$ which is a $(1+1)d$ SPT whose boundary does not admit a
unique fully symmetric ground state.

This ``decorated monopole line" construction can be generalized to
many other SPT states with mixed 1-form and 0-form symmetries. One
just need to decorate the Dirac monopole line in the $4d$ space
with a nontrivial $1d$ bosonic SPT state with ordinary 0-form
symmetry, for example the $1d$ Haldane phase with pSU($N$)
symmetry with a $\SU(N)$ fundamental at the boundary. Then the
$3d$ boundary could be described by a CP$^{N-1}$ model with $N$
flavors of the bosonic field $z_\alpha$ coupled with the dual
$\U(1)$ gauge field in Eq.~\ref{cp1}.

The system with a 0-form $\SO(3)$ symmetry and a $\U(1)^g$ 1-form
symmetry can also support other $3d$ boundary state. For example,
one can condense the bound state of a pair of the Dirac monopoles,
which can be a singlet of $\SO(3)$ 0-form symmetry. Then the
system enters a ``monopole superconductor", which is a $Z_2$
topological order with both point and loop excitations. The point
excitation is a spin-1/2 of the $\SO(3)$ 0-form symmetry
($z_\alpha$ in Eq.~\ref{cp1}), while the loop excitation carries a
half charge of the $\U(1)^g$ 1-form symmetry. This
fractionalization of 1-form symmetry is identical to the simple
fact that in an ordinary superconductor, the vortex line carries
half magnetic flux quantum. Due to the fractionalization of the
1-form symmetry, the loop excitation must couple to a gauge field,
which is precisely the 2-form gauge field dual to the condensed
Dirac monopole pair.

The $3d$ topological order constructed here is a 1-form symmetry
enriched topological (SET) states. Both the point excitation, and
the line excitation of the topological order carry nontrivial
quantum numbers, and the line excitations carry 1-form symmetry
charge. Moreover, this SET state is anomalous, in the sense that
it cannot be driven to a trivially gapped phase without breaking
either the $\U(1)$ 1-form symmetry, or the SO(3) 0-form symmetry.
The reason is that, in order to drive the $3d$ topological order
to a trivial phase, we need to either condense the point particle,
or the line excitation. However, condensing the line excitation
would lead to spontaneous breaking of the $\U(1)$ 1-form symmetry,
while condensation of the point particle would leads to
spontaneous breaking of the SO(3) symmetry.

This $Z_2$ topological order with fractionalized 1-form symmetry
is the $3d$ analogue of a $2d$ $Z_2$ topological order whose
mutual semionic anyon excitations (the so called $e$ and $m$
excitations) carry half charge and spin-1/2 representation of
0-form $\U(1)$ and $\SO(3)$ symmetries respectively. This $2d$
$Z_2$ topological order is the boundary of a $3d$ ordinary SPT
state~\cite{senthilashvin}.

The example of SET state discussed here is a prototype, namely
many 1-form SET states can be constructed in a similar way. For
example, if the $1d$ SPT state decorated in the Dirac monopole
line in the $4d$ bulk is the pSU$(N)$ Haldane phase, the $3d$
boundary can be driven into a $Z_N$ topological order by
condensing a $N$-body bound state of the Dirac monopole, which can
be a $\SU(N)$ singlet. The point particle of the $Z_N$ topological
order carries fractionalized quantum number of pSU$(N)$, and the
line excitation carries fractionalized 1-form symmetry. We leave
the full discussion of higher symmetry enriched SET states to
future exploration.

\section{3d system with tensor symmetries}

Now we a system with a generalized tensor 1-form symmetry, whose
lattice realization was discussed in
Ref.~\onlinecite{xu2006a,gu2006,xu2006b}. Connection between this
system as well as similar tensor gauge
theories~\cite{xu2008,xuhorava,xu2016} and the fracton states was
pointed out in recent literature (for instance
Ref.~\onlinecite{pretko2017a,pretko2017b,slagle2017a,slagle2017b,bulmash2018,pai2018,wang2019a,wang2019b,hermele2020,shenoy2020,seiberg2020,seiberg2020a,seiberg2020b,fontana2020,nguyen2020,gromov2019,gromov20191,gromov2020,gromov20201}).
The fracton states are a series of novel gapped states of matter,
which can be obtained by partially breaking the gauge invariance
in the generalized tensor gauge theories. In our current note we
will still focus on the gapless phase with the tensor symmetry,
instead of the gapped phase. This tensor 1-form symmetry is to
certain extent similar to three 1-form $\U(1)^g$ symmetries
discussed in the previous sections, meaning that with a given
closed surface $A$, there are three $\U(1)$ charges: $Q^a_A =
\int^A_{\partial A = \emptyset} d\vec{S} \cdot \vec{\rho}^a =
\int^V_{\partial V = A} d^3x \ \vn \cdot \vec{\rho}^a $. These
charges are individual constants. We further demand that
$\rho^{ia}$ is a symmetric tensor: $\rho^{ij} = \rho^{ji}$. Then
$\rho^{ij}$ can be viewed as the generalized symmetric tensor
electric field introduced in
Ref.~\onlinecite{xu2006a,gu2006,xu2006b}: $\cE^{ij}$, which is
subjected to the constraints: $\partial_i \cE^{ij} = \partial_j
\cE^{ij} = 0$.

Now we promote $\cE^{ij}$ to an operator $\hE^{ij}$, whose
eigenvalues are again integers. We can define the following
operator $\hat{G}(f^i(\bx))$ parameterized by an arbitrary vector
function $f^i(\bx)$: \beqn \hat{G}(f^i) &=& \exp\left( \int d^3x \
\ii 2 f^i \partial_j \hE^{ij}\right) \cr\cr &=& \exp\left( \int
d^3x \ \ii f^i \partial_j \hE^{ij} + \ii f^j \partial_i \hE^{ij}
\right) \cr\cr &=& \exp\left( - \int d^3x \ \ii \left( \partial_j
f^i + \partial_i f^j \right) \hE^{ij} \right). \eeqn Let us denote
$\hA^{ij}$ as the canonical conjugate operator of $\hE^{ij}$
($\hA^{ij}$ is again periodically defined). More precisely, we
impose the commutation relations $[\hE^{ij}(\bx),
\hA^{i'j'}(\bx')]= \ii (\delta_{ii'} \delta_{jj'}+\delta_{ij'}
\delta_{ji'}) \delta(\bx-\bx')$. The $\hat{G}(f^i)$ operator will
generate a gauge transformation on $\hA^{ij}$: \beqn
&\hat{G}^{-1}(f^i) \hA^{ij}(\bx) \hat{G}(f^i) = \hA^{ij}(\bx) +
2\partial_i f^j + 2\partial_j f^i, ~~~~ \label{gauge} \eeqn
However, because of the constraint on $\hE^{ij}$, $\hat{G}(f^i)$
is actually an identity operator, which must commute with any
Hamiltonian of $\hE^{ij}$ and $\hA^{ij}$. It means that the
Hamiltonian of the system {\it must} be invariant under the gauge
transformation Eq.~\ref{gauge}. The derivation of gauge invariance
in this paragraph applies to other systems with local constraints,
such as systems with generalized gauge
transformations~\cite{xu2016}.

Then the Hamiltonian must be a function of $\hE^{ij}$ and the
gauge invariant operator $\hB^{ij} =
\epsilon_{iab}\epsilon_{jcd}\partial_a \partial_c \hA^{bd}$. A
general local Hamiltonian should take the form: \beqn H =
\sum_{\bx} \cH [\hE^{ij}(\bx), \ \hB^{ij}(\bx)], \eeqn and again
$\cH$ is a periodic function of $\hB^{ij}$. $\hB$ is completely
dual to $\hE$. Besides the more exotic gauge invariance, these
Hamiltonians all have an extra center of mass conservation: $H$ is
invariant under transformation \beqn \hA^{ij} \rightarrow \hA^{ij}
+ F^{ij}[\bx], \label{com} \eeqn where $F^{ij}[\bx]$ is a linear
function of space coordinate. This extra conservation law in the
series of tensor models~\cite{xu2006a,xu2006b,xu2016} was noticed
in Ref.~\onlinecite{pretko2017a}, and it was realized that this
center of mass conservation is a key feature of the fracton states
of matter.

We can define a dual tensor 1-form symmetry $\U(1)^g_\dual$, whose
charge corresponds to the generalized tensor magnetic flux through
a surface $A$: $Q^{a}_{A} = \int dS^i \cdot \hB^{ia}$. We assume
that the generalized tensor magnetic 1-form charge density
$\partial_i \hB^{ij} = \partial_j \hB^{ij} = 0$ remains zero at
low energy, meaning there is an emergent dual tensor symmetry
$\U(1)^g_\dual$ at long scale. Then again one can insert magnetic
flux through the system through (for example) the following
operator: \beqn \hat{O}_{q} = \exp\left(\ii q \sum_{\bx}
\frac{2\pi x^2}{L^2} \hE^{zz}(\bx) \right). \label{Oq2}\eeqn This
operator is still compatible with the periodic boundary condition,
and it will shift $\cA^{zz}$ by \beqn
\hat{O}^{-1}_{q}\hA^{zz}\hat{O}_{q} = \hA^{zz}(\bx) + \frac{4\pi q
x^2}{L^2}.\eeqn If we denote the ground state of the system as
$|\Omega\rangle$, then $|\Psi\rangle$ has nonzero extra quantized
flux of $\cB^{yy}$ through any XZ plain compared with the ground
state, and the extra flux density is $\cB^{yy} \sim 1/L^2 $. Or we
can create a configuration of $\cA^{xy}(\bx)$ as $ \cA^{xy}(\bx) =
2\pi z^2/L^2 $. Then there is a nonzero flux of $\cB^{xy}$, again
with flux density $\sim 1/L^2$.

Again we will demonstrate that the ground state of the system
cannot be trivially gapped, if we assume the emergent $\U(1)^g
\times \U(1)^g_\dual$ symmetry at long scale. Suppose there is a
unique ground state $|\Omega\rangle$ of the system, then
$|\Psi\rangle = \hat{O}_q |\Omega\rangle$ must be orthogonal to
$|\Omega\rangle$ for large enough $L$, because $|\Omega\rangle$
must be an eigenstate of the tensor 1-form charge, and
$|\Psi\rangle$ carries a different tensor 1-form charge from
$|\Omega\rangle$. And by going through the same argument as
section II, we can demonstrate that when $L \rightarrow \infty$,
the energy of $|\Psi\rangle$ must also approach the energy of
$|\Omega \rangle$. This statement still holds with disorder, and
also when there is a long range interaction that falls off more
rapidly than $1/|\bx|^2$.

We have argued that an emergent $\U(1)^g \times \U(1)^g_\dual$
tensor symmetry rules out a trivial gapped ground state. This
result can be equivalently stated as that the $\U(1)^g \times
\U(1)^g_\dual$ tensor symmetry is anomalous. Again, the equation
of motion of $\cE^{ij}$ and $\cB^{ij}$ can be viewed as the
continuity equation of the currents of the tensor symmetries. For
the simplest semiclassical limit of the
theory~\cite{xu2006a,xu2006b}, the Hamiltonian of the system is
approximately \beqn \cH \sim \frac{1}{4} \sum_{\bx}\sum_{i j}
\left[  \ \left(\cE^{ij}(\bx) \right)^2 +
\left(\cB^{ij}(\bx)\right)^2 \right], \eeqn then the equation of
motion reads \beqn \frac{\partial \cE^{ij}}{\partial t} -
\partial_a \left(\epsilon_{iab} \partial_c \epsilon_{jcd} \cB^{bd} \right)
= 0, \cr\cr \frac{\partial \cB^{ij}}{\partial t} - \partial_a
\left(\epsilon_{iab} \partial_c \epsilon_{jcd} \cE^{bd} \right) =
0. \eeqn This means that the currents of the tensor symmetries
are: \beqn J^{(e)} = (\rho^{(e)}_{ij}, \ J^{(e)}_{ij,k}) =
\left(\cE^{ij}, \ \frac{1}{2}\epsilon_{ikb} \epsilon_{jcd}
\partial_c \cB^{bd} + i \leftrightarrow j \right), \cr\cr J^{(m)}
= (\rho^{(m)}_{ij}, \ J^{(m)}_{ij,k}) = \left(\cB^{ij}, \
\frac{1}{2}\epsilon_{ikb} \epsilon_{jcd} \partial_c \cE^{bd} + i
\leftrightarrow j \right). \label{emTensorCurrent} \eeqn Again in
a process that creates a nonzero current of one of the $\U(1)$
tensor symmetries, the charge conservation of the other $\U(1)$
tensor symmetry must be violated, hence there is a 't Hooft
anomaly of the two $\U(1)$ tensor symmetries.

Formally we can still discuss the anomalies in a Lagrangian
formalism. The Lagrangian is given by
\begin{align}
\mathcal{L}[\dot{\cA}^{ij}, \cB^{ij}] = \frac{1}{4} \sum_{\bx}
\sum_{ij} \left[ \left( \dot{\cA}^{ij} (\bx)\right)^2 - \left(
\cB^{ij} (\bx) \right)^2 \right],
\end{align}
where $\dot{\cA}^{ij}  \equiv \delta H/ \delta \cE^{ij}= \cE^{ij}$
is introduced through the Legendre transformation $\mathcal{L} =
\left(\sum_{\bx} \sum_{ij}\cE^{ij} \frac{\delta H}{\delta
\cE^{ij}} \right) - H$. The electric $\U(1)^g$ tensor symmetry is
defined by the symmetry transformation
\begin{align}
\cA^{ij} \rightarrow \cA^{ij}+ \Lambda^{(e)}_{ij},
\end{align}
where $\Lambda^{(e)}_{ij}$ is a constant symmetric tensor, namely
$\Lambda^{(e)}_{ij} = \Lambda^{(e)}_{ji}$. In the following, we
will use the terms $\U(1)^g$ tensor symmetry and electric tensor
symmetry interchangeably. We can gauge the electric tensor
symmetry by promoting $\Lambda^{(e)}_{ij}$ to a space-time
function, and introducing the electric tensor gauge fields
$G^{(e)}_{ij,0}$ and $G^{(e)}_{ij,k}$ which are symmetric under
the exchange of the first two indices, namely
$G^{(e)}_{ij,0}=G^{(e)}_{ji,0}$ and
$G^{(e)}_{ij,k}=G^{(e)}_{ji,k}$. Under the electric tensor gauge
transformation, we have
\begin{align}
& \cA^{ij} \rightarrow \cA^{ij}+ \Lambda^{(e)}_{ij} \nn
\\
& \dot{\cA}^{ij}  \rightarrow  \dot{\cA}^{ij} +
\partial_t\Lambda^{(e)}_{ij} \nn
\\
& \cB^{ij} \rightarrow \cB^{ij} + \epsilon_{iab}\epsilon_{jcd}
\partial_a \partial_c \Lambda^{(e)}_{bd} \label{eTensorGaugeTrans}
\\
& G^{(e)}_{ij,0} \rightarrow G^{(e)}_{ij,0} + \partial_t
\Lambda^{(e)}_{ij} \nn
\\
& G^{(e)}_{ij,k} \rightarrow  G^{(e)}_{ij,k} + \partial_k
\Lambda^{(e)}_{ij} \nn,
\end{align}
When the electric tensor background gauge field is turned on, the
system is described by the Lagrangian
\begin{align}
& \mathcal{L}\left[\dot{\cA}^{ij} - G^{(e)}_{ij,0} , \cB^{ij} -
\frac{1}{2} \left(\epsilon_{iab} \epsilon_{jcd} + \epsilon_{icb}
\epsilon_{jad} \right) \partial_a  G^{(e)}_{bd,c} \right]
\nn \\
& = \mathcal{L}[\dot{\cA}^{ij}, \cB^{ij} ] - \sum_{ijk} \sum_{\bx}
\left( G^{(e)}_{ij,0} \rho_{ij}^{(e)} + G^{(e)}_{ij,k}
J_{ij,k}^{(e)} \right)+...,
\end{align}
which is explicitly gauge invariant under the gauge transformation
given by Eq.~\ref{eTensorGaugeTrans}. The ``..." part contains
higher order terms in $\dot{\cA}^{ij}$ and $\cB^{ij}$. As a sanity
check, we notice that the Lagrangian above effectively introduces
the minimal coupling between the electric tensor gauge fields
$\left( G^{(e)}_{ij,0}, ~G^{(e)}_{ij,k}\right) $ and the current
$J^{(e)}$ introduced in Eq.~\ref{emTensorCurrent}.

Similarly, we can introduce the magnetic tensor gauge fields
$G^{(m)}_{ij,0}$ and $G^{(m)}_{ij,k}$ which are also symmetric
under the exchange of the first two indices, namely
$G^{(m)}_{ij,0}=G^{(m)}_{ji,0}$ and
$G^{(m)}_{ij,k}=G^{(m)}_{ji,k}$. The magnetic tensor gauge fields
are associated to the emergent $\U(1)^g_{\rm dual}$ symmetry. They
transform under the magnetic tensor gauge transformation as
\begin{align}
& G^{(m)}_{ij,0} \rightarrow G^{(m)}_{ij,0} + \partial_t
\Lambda^{(m)}_{ij}, \nn
\\
& G^{(m)}_{ij,k} \rightarrow  G^{(m)}_{ij,k} + \partial_k
\Lambda^{(m)}_{ij}.
\end{align}
We can introduce the minimal coupling between the magnetic tensor
gauge fields and the current $J^{(m)}$ introduced in Eq.
\ref{emTensorCurrent}, which yields
\begin{align}
& \mathcal{L}_g  =  \mathcal{L}\left[\dot{\cA}^{ij} -
G^{(e)}_{ij,0} , ~\cB^{ij} - \frac{1}{2} \left(\epsilon_{iab}
\epsilon_{jcd} + \epsilon_{icb} \epsilon_{jad} \right) \partial_a
G^{(e)}_{bd,c} \right]
\nn \\
& ~~~ - \sum_{ijk} \sum_{\bx}\left(  G^{(m)}_{ij,0}
\rho_{ij}^{(m)} + G^{(m)}_{ij,k} J_{ij,k}^{(m)}  \right) \nn
\\
& = \mathcal{L}\left[\dot{\cA}^{ij} - G^{(e)}_{ij,0} , ~\cB^{ij} -
\frac{1}{2} \left(\epsilon_{iab} \epsilon_{jcd} + \epsilon_{icb}
\epsilon_{jad}\right) \partial_a  G^{(e)}_{bd,c} \right]
\nn \\
&  - \sum_{ijk} \sum_{\bx}\left(  G^{(m)}_{ij,0} \cB^{ij} +
\frac{1}{2}G^{(m)}_{ij,k} \left( \epsilon_{ikb} \epsilon_{jcd} +
\epsilon_{jkb} \epsilon_{icd}\right)
 \partial_c \dot{\cA}^{bd}  \right).
\end{align}
When the magnetic tensor gauge field $\left( G^{(m)}_{ij,0}, \,
G^{(m)}_{ij,k} \right)$ is turned on, the Lagrangian
$\mathcal{L}_g$ is no longer invariant under the electric tensor
gauge transformation Eq. \ref{eTensorGaugeTrans}:
\begin{align}
\mathcal{L}_g \rightarrow \mathcal{L}_g - & \sum_{ijk} \sum_{\bx}
\left( G^{(m)}_{ij,0} \epsilon_{iab}\epsilon_{jcd} \partial_a
\partial_c \Lambda^{(e)}_{bd} \right.
\nn \\
& \left. + \frac{1}{2}G^{(m)}_{ij,k} \left( \epsilon_{ikb}
\epsilon_{jcd} + \epsilon_{jkb} \epsilon_{icd}\right)
 \partial_c \partial_t \Lambda^{(e)}_{bd}
\right).
\end{align}
The fact that $\mathcal{L}_g$ is no longer gauge invariant once
the magnetic tensor gauge field $\left( G^{(m)}_{ij,0}, \,
G^{(m)}_{ij,k} \right)$ is turned on indicates a 't Hooft anomaly
of the emergent $\U(1)^g \times \U(1)^g_\dual$ tensor symmetry.

The $3d$ system with tensor symmetry can also be compactified to
$1d$. After compactification, one can still define several
ordinary $1d$ global $\U(1)$ symmetries. One of the $\U(1)$
symmetries has the following charge: \beqn \int_{\bx \in
\mathrm{XY}} d^2x \ \hE_{zz}(\bx) \sim \int dx \ \hn(x). \eeqn The
conjugate variable of $\hn(x)$, i.e. the phase angle $\htheta(x)$
is defined as \beqn \int_{\bx  \in \mathrm{XZ}} d^2x \
\hB_{yy}(\bx) \sim \int dx \ \nabla_x^2 \htheta(x), \eeqn and
$\htheta(x) \sim \hA_{zz}(x)$. The $3d$ Hamiltonian then reduces
to a $1d$ Lifshitz theory: $H = \sum_x \cH[\hn(x), \ \nabla_x^2
\htheta(x)]$.

The $1d$ Hamiltonian $H$ inherits the center of mass conservation
Eq.~\ref{com}, which in $1d$ becomes $\htheta \rightarrow \htheta
+ B x$ with constant $B$. This center of mass conservation
prohibits terms like $\cos(\nabla_x \htheta)$ after
compactification. Hence after compactification, the $2d$ bulk of
the system should be an exotic SPT state with a special center of
mass conservation, whose nature deserves further studies.

\section{Discussion}

In this note we explored the consequences of the assumption of a
pair of dual generalized symmetries. We discussed the implication
of the dual symmetries on low energy excitations, 't Hooft
anomaly, their bulk description, and corresponding state after
dimensional compactification. Then we extended all these
discussions to the tensor gauge theories.

Further studies can be pursued following the questions raised in
this work. We have shown that, for $N$ pairs of dual 1-form
symmetries in $3d$, there will be $2N$ pairs of dual 0-form
symmetries after compactification to $1d$. If we break the dual
1-form symmetries to certain combination of these two 1-form
symmetries, a bound state of electric and magnetic charges (a
dyon) is allowed and has its own dynamics. The $3d$ system can be
driven to a gapped phase by condensing these dyons, and the gapped
$3d$ system may have a topological order which depends on the
condensed object. There should be a systematic formalism
describing the relation between the gapped $3d$ systems and the
corresponding gapped $1d$ systems after dimensional
compactification. The problem would be further enriched if there
is topological $\Theta-$term in the $3d$
system~\cite{cardy1982a,cardy1982b}.

The $1d$ system after compactification is described by ordinary
boson operators $\hat{n}$ and $\htheta$, and these bosons do not
fractionalize. Hence it is sufficient to view the $1d$ system as
the boundary of a $2d$ SPT state, instead of a $2d$ topological
order with fractionalization. The $4d$ bulk of the $3d$ system is
also a generalized SPT state with 1-form symmetries, rather than a
topological order. But a topological order with fractionalized
1-form symmetries would be an interesting direction to explore. In
section IV we presented one prototype of such topological order. A
more general and systematic discussion of fractionalization of
higher form symmetry is worth studying in the future.

Besides the higher-form symmetries and tensor like symmetries,
many other generalized concepts of symmetries have been discussed
in the past (for early examples please see
Ref.~\onlinecite{paramekanti,nussinov1,xufisher,nussinov2}). Much
of the topics discussed in this paper, such as the SPT states and
anomalies involving these generalized symmetries are also
interesting future directions.

This work is supported by NSF Grant No. DMR-1920434, the David and
Lucile Packard Foundation, and the Simons Foundation.

\bibliography{anomaly}

\end{document}